\documentclass[aps,psfig,showpacs,groupedaddress]{revtex4}
\usepackage{color}
\usepackage{graphics}
\usepackage{epsfig}
\usepackage{ulem}
\usepackage{amsmath,environ}

\newcommand{\bee}{\begin{equation}}
\newcommand{\ene}{\end{equation}}
\newcommand{\beea}{\begin{eqnarray}}
\newcommand{\enea}{\end{eqnarray}}

\begin{document}
\title{Energy principle  for 2D electromagnetic, relativistic, interpenetrating, counterstreaming plasma flows}
\author{Atul Kumar}
\email{atul.kumar@ipr.res.in}
\author{Predhiman Kaw} 
\author{Amita Das}
\email{amita@ipr.res.in}

\affiliation{Institute for Plasma Research, HBNI,  Bhat, Gandhinagar - 382428, India }

\begin{abstract} 
A relativistic electron beam propagating through plasma induces a return current in the system. Such a system of counterstreaming forward and return current is susceptible to host of instabilities out of which Weibel remains a dominant mode for destabilizing the system. Weibel instability has been widely investigated in simulations, experiments, as well as, analytically using fluid and kinetic treatments.  A purely growing mode like Weibel instability can also be understood by using conservation theorems and energy principle analysis. An electrostatic analog to the Weibel instability, two stream instability in a beam plasma system has already been investigated using energy principle analysis \cite{Lashmore-Davies2007}.  A detailed analytical description of conservation theorem for 2D Weibel 
instability in a beam plasma system has been carried out in this manuscript.  

\end{abstract}
\maketitle 

\section{Introduction}
An intense laser interacting with an overdense solid target generates highly energetic, relativistic electrons \cite{Modena, Brunela,PhysRevLett.81.822} carrying a very large current ($ \sim 
MAmps $) in forward direction. In response to this, the background supplies a return current. 
The spatially overlapping  forward and return current flow is susceptible to  several microinstabilities.  The current filamentation instability \cite{Pegoraro, Bret2005, Bret2010, 
Fox2013}, 
which is often referred to as the Weibel instability \citep{Weibel},  dominates over all the instabilities in the 
relativistic regime. 
 The Weibel instability creates a space charge separation of the order of electron skin depth, $ d_e $ in the plasma, which leads to a generation of giant magnetic field 
( $ \sim MGauss $ ) through a positive feedback mechanism. It is  believed that the Weibel instability and its nonlinear evolution is largely responsible 
for the development of giant magnetic fields in 
 astrophysical contexts like the relativistic shock formation in Gamma ray burst mechanisms(GRBs) \cite{Mirabel}, the high cosmic rays \cite{Bell, Ackermann}, active galactic nucleii (AGN)  \cite{Bridle} etc.  In numerous laser- plasma 
 laboratory experiments  the Weibel instability driven by two counter-streaming high-energy flows has also been  demonstrated\citep{Mondal, Huntington2015}.\\

The study of linear and nonlinear evolution of giant magnetic fields through  Weibel destablization process is,  therefore, of prime importance for the understanding of various 
astrophysical events as well as laboratory experiments. Califano {\it{et. al.}} \cite{PhysRevE.58.7837},  have  given an analytical model in 2D for  relativistic, interpenetrating, 
homogeneous beam-plasma system based on two fluid depiction  for this particular instability. In a typical beam plasma system, when  a minuscule fraction of 
  kinetic energy ( $ \delta K$) associated with the flow 
gets  converted into electromagnetic excitations it essentially leads to the excitation of negative energy disturbances, which feed upon themselves leading to the 
development of the instability. The instability is then responsible for the generation of giant magnetic fields in the system. \\

Energy principle is a well known technique for the stability analysis of hydrodynamic fluids in a conservative system \cite{Bernstein17,Frieman_1960,Kulsrud,Isichenko_1998,Hameiri_1998,Davidson_2000}.  In plasmas also the ideal Magnetohydrodynamic (MHD) 
excitations have often been interpreted using the energy principle \cite{Laval_1965,Freidberg_1987, Hameiri_2003,Khalzov}. This has been done for both kinds of equilibria, namely  with and without flows . 
 We often encounter zero frequency mode  in ideal MHD where two modes couple to lead a purely growing  zero frequency excitation. 
 The stability theory with energy principle have been aplplied for ideal magnetohydrodynamics in a  variety of contexts like magnetic  fusion, astrophysics, solar and space physics 
 etc. \\

Weibel instability is an instability of mixed electrostatic and electromagnetic waves with latter playing a crucial role 
when the wavenumber  is transverse to the flow direction corresponding to  current filamentation. In this case it  is purely growing zero frequency mode.   
We show here that the energy principle can be used to explain the growth of this particular mode. 
In an earlier  study 
Lashmore-Davis \cite{Lashmore-Davies2007} has  shown the development of two 
stream electrostatic instability using  energy principle for a beam plasma system.   In this paper, we employ the  energy principle analysis for the development of 
  the $ 2D $ electromagnetic, relativistic, homogeneous beam plasma system in which the beam and plasma return currents are spatially overlapping. \\

The manuscript has been organized as follows. Section II describes the system and contains the derivation of the model set of equations for  analysis.
In section III the detailed analysis for the specific case of  current filamentation/Weibel instability  has been carried out. 
Section IV contains the summary. 

\section{Governing Equations}
In the presence of plasma medium the Maxwell's equations  
 \begin{eqnarray}
 \vec{\nabla} \times \vec{B} = \frac{4 \pi}{c} \vec{J}+ \frac{1}{c} \frac{\partial \vec{E}}{\partial t};  
 \nonumber \\
 \vec{\nabla} \times \vec{E} = -\frac{\partial \vec{B}}{\partial t};
\label{Maxwell}
\end{eqnarray}
can be used to obtain the following equation for the evolution of field energy 
\begin{equation}
{\frac{\partial}{\partial t}\bigg( \frac{E^2+B^2}{8\pi}\bigg)  + \frac{c}{4\pi}\vec{\nabla}\cdot (\vec{E} \times \vec{B}) + \vec{J} \cdot \vec{E} = 0}
\label{conser}
\end{equation}
In vacuum the $\vec{J} \cdot \vec{E}$ term is absent and the rate of change of electromagnetic energy is determined by the Poynting flux. In plasma or any  conducting media
 the currents can flow so as to have $\vec{J} \cdot \vec{E}$  finite. This term essentially represents the possibility of energy transfer between the kinetic energy of the particles to 
 field energy and vice versa. 
 
\begin{figure}[h!]
\center
                \includegraphics[width=\textwidth]{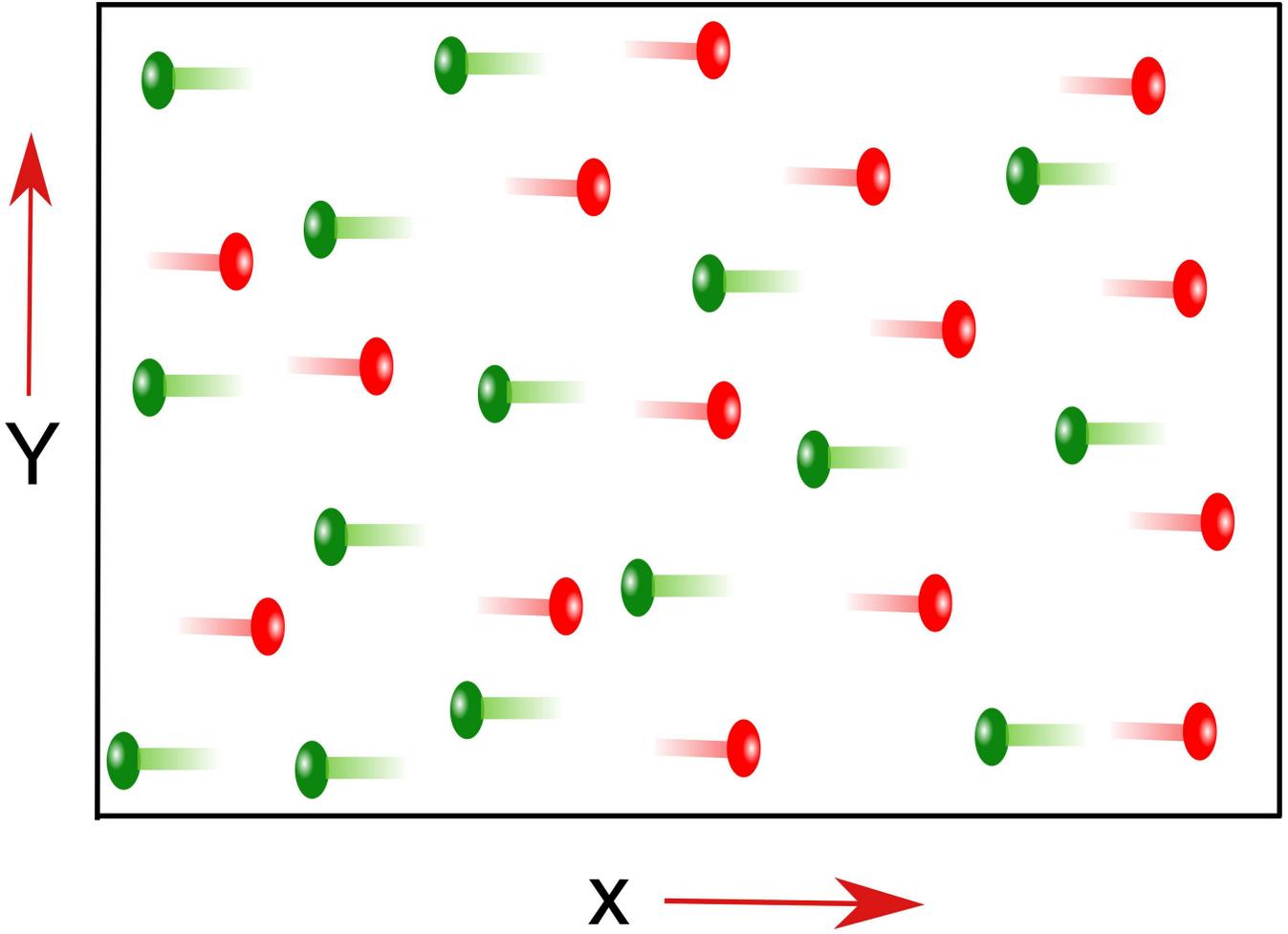} 
             \caption{ Schematics of 2D- equilibrium geometry of the beam plasma system   }  
                 \label{fig1}
         \end{figure}

For an equilibrium  configuration in the plasma the time dependent electric and magnetic field fluctuations are absent. The fluctuations excited around a homogeneous 
equilibrium can be represented by a collection of Fourier modes having variations in the fields of the form 
$f(\vec{r},t) \sim f_{k,\omega} exp(i \vec{k} \cdot \vec{r} - i \omega t) + c.c $, where $c.c$ represents the complex conjugate. The real part of 
$ \vec{J}\cdot \vec{E} $ would thus have the contribution from 
\begin{equation}
{\vec{J} \cdot \vec{E} = \frac{1}{2}(\vec{J}\cdot \vec{E}^* + \vec{J}^* \cdot \vec{E})}
\label{current}
\end{equation}
We now consider a system depicted in  Fig.(\ref{fig1}). A rectangular box in $x-y$ plane has been shown in which 
the beam (red circles ) and background plasma electrons (green circles) are flowing in positive and negative $x$ directions respectively. We will represent the beam and background plasma electrons 
by suffix $b$ and $p$ respectively. An equilibrium configuration is chosen in which the charge of the 
beam and plasma electron densities are neutralized by the background plasma ions. We thus have 
$\sum_{\alpha}n_{0\alpha} = n_{0i}$, here $\alpha$ is a dummy index which is equal to $b$ and $p$ 
to represent the beam and plasma electrons respectively, $n_{0\alpha}$ is the density 
of the two species and $n_{0i}$ is the background ion density. 
We also assume a flow of beam and 
background plasma electrons in equilibrium in opposite directions so as to have 
the total electron current in the system to be zero, i.e.  
 $ \sum_{\alpha} n_{o\alpha	} \vec{v}_{0\alpha} =0 $, where $\vec{v}_{0\alpha}$ is chosen 
 to be the equilinrium flow velocity of the two species. We will assume the flow 
 velocities to be directed along the positive and negative $x$ axis as shown in Fig.1 for the beam and 
 background plasma electrons.  The charge and current neutrality of the system ensures the absence of 
 any electric and magnetic fields. We assume such a counter streaming plasma to be of infinite extent 
 by considering periodic boundary condition along $x$ and $y$ directions. 
 
 The linearized Weibel excitations for such an equilibrium configuration has been considered. The Weibel 
 excitations have a mixed electrostatic and electromagnetic character. The variations are assumed to 
 be along the $y$ directions only. Thus the wave vector $k$ is directed along $y$. It 
 is easy to see that such a configuration leads to perturbed magnetic field which is directed along 
 $\hat{z}$ and is denoted by $B_{1z}$.   The perturbed electric field $ \vec{E}_1 $ lies in the x-y 
 plane. The variations being along $\hat{y}$, $E_{1y}$ is the electrostatic 
component of the field and $ E_{1x} $, $ B_{1z} $ correspond to  electromagnetic excitation. 
These perturbations thus convert the flow kinetic energy of the system into field energies. Since 
the medium is infinite in extent there is no loss or gain of energy by the Poynting flux. 
The energy conversion occurs through the 
 $ \vec{J} \cdot \vec{E} $ term which represents 
  the work done by or on the particles and needs to be evaluated for the perturbed excitations. 
  Keeping in view the quadratic nature of energy one retains all linearized contributions for 
  $\vec{J}$ and $\vec{E}$ separately in the evaluation and try to express  $ \vec{J} \cdot \vec{E} $ 
  in a quadratic form. 
  This implies that  
\begin{equation}
{\vec{J} \cdot \vec{E}^* = -\sum \bigg[e n_{1\alpha} v_{0\alpha x} E_{1x}^* +n_{0\alpha} v_{1\alpha x} E_{1x}^* + en_{0\alpha} v_{1\alpha y}E_{1y}^*\bigg]  }
\label{linearJE}
\end{equation}
The perturbed density and velocity contributions for the two species can be obtained from the linearized 
continuity and momentum equations of the two species.
\begin{equation}
{\frac{\partial n_{1\alpha}}{\partial t} + \frac{\partial}{\partial y} (n_{0\alpha}v_{1\alpha y})} =0;
\label{contEq}
\end{equation} 
 \begin{eqnarray}
 {\frac{\partial}{\partial t} (\gamma_{0\alpha}^3 v_{1\alpha x}) = eE_{1x}}; \hspace{0.2in}
{\frac{\partial}{\partial t} (\gamma_{0\alpha}v_{1\alpha y}) = eE_{1y}+ ev_{0\alpha x}B_{1z}};
\label{MomEq}
\end{eqnarray}
Now taking Fourier transform in time $t$ and space  $y$  Eq.(\ref{contEq},\ref{MomEq}) and the Maxwell's  Eq.(\ref{Maxwell}) 
and using $ m_e=1 $, $ c=1$, can be written as 
 \begin{eqnarray}
 n_{1\alpha} &=& \frac{k}{\omega} n_{0\alpha} v_{1\alpha y} \nonumber \\
v_{1\alpha x} &=& -\frac{\iota e}{\gamma_{0\alpha}^3 \omega}E_{1x}  \nonumber \\
v_{1\alpha y} &=& -\frac{\iota e}{\gamma_{0\alpha} \omega}E_{1y} - \frac{\iota e}{\gamma_{0\alpha}\omega^2} kv_{0\alpha x} E_{1x}    \nonumber \\
B_{1z} &=& -\frac{k}{\omega} E_{1x} 
\label{linEqn}
\end{eqnarray}
In addition 
the  fourier transform of the Poisson's equation upon
 substituting for  $ n_{1\alpha} $ from  Eq.(\ref{linEqn}) can be written as 
\begin{equation}
{ \iota k E_{1y} = - \sum_{\alpha} 4 \pi e \frac{k}{\omega} n_{0\alpha} \bigg[ -\frac{\iota e}{\gamma_{0\alpha} \omega}E_{1y} - \frac{\iota e}{\gamma_{0\alpha}\omega^2} kv_{0\alpha x} E_{1x} \bigg] }
\label{linPoisson}
\end{equation}
We will now use these relationships to eliminate all variables (e.g. $n_{1\alpha}, \vec{v}_{1\alpha}, E_{1y}, B_{1z}$) in terms of $E_{1x}$ and express the 
energy equation in terms of $E_{1x}$ variable alone. 
Defining 
\begin{eqnarray}
S_1 &=& \sum_{\alpha} \frac{n_{0 \alpha}}{n_0 \gamma_{0 \alpha}}; \\
S_2 &=& \sum_{\alpha} \frac{n_{0 \alpha}}{n_0 \gamma_{0 \alpha}^3};\\
S_3 &=& \sum_{\alpha} \frac{n_{0 \alpha} v_{0 x \alpha}}{n_0 \gamma_{0 \alpha}}; \\
S_4 &=& \sum_{\alpha} \frac{n_{0 \alpha} v_{0 x \alpha}^2}{n_0 \gamma_{0 \alpha}};
\label{sdefs}
\end{eqnarray}
The Poisson's  equation  (Eq.(\ref{linPoisson})) relates the $x$ and $y $ components of the electric field by the following expression 
\begin{equation}
{ E_{1y} = \frac{k S_3}{(1 - \frac{\omega_{pe}^2}{\omega^2}S_1) \omega^3} E_{1x} }
\label{PoissonEqn}
\end{equation}
This relationship Eq.(\ref{PoissonEqn}) is utilized to express other fields in terms of $E_{1x}$
as 
\begin{equation}
{ v_{1\alpha x} = -\frac{\iota e}{\gamma_{0\alpha}^3 \omega}E_{1x} }
\label{v1alphax}
\end{equation}
\begin{equation}
{ v_{1\alpha y} = - \iota \bigg[ \frac{\omega_{p}^2 S_3 k e}{\omega^4 (1- \frac{S_1}{\omega^2})\gamma_{0\alpha} } E_{1x} + \frac{ekv_{ 0 \alpha x}}{\omega^2 \gamma_{0\alpha}}E_{1x} \bigg] }
\label{v1alphay}
\end{equation}
\begin{equation}
{ n_{1\alpha } = - \iota \bigg[ \frac{\omega_p^2 	S_3 k^2 e n_{0 \alpha}}{\omega^5 (1- \frac{S_1}{\omega^2})\gamma_{0\alpha} } E_{1x} + \frac{ek^2 v_{ 0 \alpha x}n_{0 \alpha}}{\omega^3 \gamma_{0\alpha}}E_{1x} \bigg] }
\label{n1alpha}
\end{equation}
 
\section{Energy Conservation theorem for 2D Weibel Instability}
For the infinite counterstreaming beam and background plasma electrons, the Poynting flux  would be zero and hence can be neglected from 
 Eq.(\ref{conser})  which takes the form 
 \begin{equation}
{\frac{\partial}{\partial t}\bigg( \frac{E^2+B^2}{8\pi}\bigg) + \frac{1}{2}\bigg[\vec{J} \cdot \vec{E}^* + \vec{J}^*\cdot\vec{E} \bigg] = 0}
\label{conserhomo}
\end{equation}
We write  $ \vec{J} \cdot \vec{E} $ given in Eq.(\ref{linearJE}) upon retaining $ 2^{nd} $ order perturbations  as,
\begin{equation}
{ \vec{J} \cdot \vec{E} = \frac{1}{2} [(A+A^*) + (B+ B^*) + (C+C^*)]  }
\end{equation} 
Here  A, B and C are given by 
\begin{eqnarray}
A &=&  - \sum_{\alpha} en_{1\alpha}v_{0\alpha x} E_{1x}^*; \\
B &=&  - \sum_{\alpha} en_{0\alpha}v_{1\alpha x} E_{1x}^*; \\
C &=&  - \sum_{\alpha} en_{0\alpha}v_{1\alpha y} E_{1y}^*; 
\label{ABC}
\end{eqnarray}
Now A, B, C and their complex conjugate expressions can be expressed entirely in terms of $E_{1x}$ and its conjugate as 
\begin{eqnarray}
{ A = \frac{\iota}{4\pi} \bigg[\frac{\omega_p^2k^2}{\omega^5(1- \frac{\omega_p^2}{\omega^2}S_1)}S_3 \sum_{\alpha} \frac{n_{0\alpha}v_{0\alpha x}}{n_0 \gamma_{0\alpha}} + \frac{\omega_p^2 k^2}{\omega^3}\sum_{\alpha} \frac{n_{0\alpha}v_{0\alpha x}^2}{n_0 \gamma_{0\alpha}} \bigg] \vert E_{1x}\vert^2 };\\
{ A^* = -\frac{\iota}{4\pi} \bigg[\frac{\omega_p^2k^2}{\omega_*^5(1- \frac{\omega_p^2}{\omega_*^2}S_1)}S_3 \sum_{\alpha} \frac{n_{0\alpha}v_{0\alpha x}}{n_0 \gamma_{0\alpha}} + \frac{\omega_p^2 k^2}{\omega_*^3}\sum_{\alpha} \frac{n_{0\alpha}v_{0\alpha x}^2}{n_0 \gamma_{0\alpha}} \bigg] \vert E_{1x}\vert^2 };\\
{B=\frac{\iota}{4\pi}\bigg[ \frac{\omega_p^2}{\omega} \sum_{\alpha} \frac{n_{0\alpha}}{n_0 \gamma_{0\alpha}^3}\bigg] \vert E_{1x} \vert ^2 };\\
{B^*= -\frac{\iota}{4\pi}\bigg[ \frac{\omega_p^2}{\omega_*} \sum_{\alpha} \frac{n_{0\alpha}}{n_0 \gamma_{0\alpha}^3}\bigg] \vert E_{1x} \vert ^2 };\\
{ C=\frac{\iota}{4\pi} \bigg[ \frac{\omega_p^6 k^2 S_3^2}{\vert \omega\vert^6 \omega (1- \frac{\omega_p^2}{\omega^2}S_1)(1- \frac{\omega_p^2}{\omega_*^2}S_1)}  \sum_{\alpha} \frac{n_{0\alpha}}{n_0 \gamma_{0\alpha} } + \frac{\omega_p^4 k^2 S_3}{\vert \omega \vert \omega_* (1- \frac{\omega_p^2}{\omega^2}S_1)} \sum_{\alpha} \frac{n_{0\alpha}v_{0\alpha x}}{n_0 \gamma_{0\alpha} }\bigg] \vert E_{1x} \vert ^2};\\
{ C^*=-\frac{\iota}{4\pi} \bigg[ \frac{\omega_p^6 k^2 S_3^2}{\vert \omega\vert^6 \omega_* (1- \frac{\omega_p^2}{\omega_*^2}S_1)(1- \frac{\omega_p^2}{\omega^2}S_1)}  \sum_{\alpha} \frac{n_{0\alpha}}{n_0 \gamma_{0\alpha} } + \frac{\omega_p^4 k^2 S_3}{\vert \omega \vert \omega (1- \frac{\omega_p^2}{\omega^2}S_1)} \sum_{\alpha} \frac{n_{0\alpha}v_{0\alpha x}}{n_0 \gamma_{0\alpha} }\bigg] \vert E_{1x} \vert ^2};
\label{ABCfinal}
\end{eqnarray}
Collecting all the terms 
and using the normalization $ \omega \equiv \omega/\omega_p $ and $ k \equiv kc/\omega_p $, the expression for $ \vec{J}\cdot \vec{E} $ can be written as, 
\begin{equation}
\begin{split}
\vec{J} \cdot \vec{E} &=\frac{\iota}{8\pi}\bigg[S_3^2 k^2 \bigg\lbrace \frac{1}{\omega^5(1- \frac{1}{\omega^2}S_1)} - \frac{1}{\omega_*^5(1- \frac{1}{\omega_*^2}S_1)} \bigg \rbrace + S_4 k^2 \bigg \lbrace \frac{1}{\omega^3} - \frac{1}{\omega_*^3}\bigg \rbrace \bigg]\vert E_{1x} \vert ^2 \\
& + \frac{\iota}{8\pi}\bigg[ S_2 \bigg \lbrace \frac{1}{\omega} - \frac{1}{\omega_*}  \bigg \rbrace \bigg]\vert E_{1x} \vert ^2\\
& +  \frac{\iota}{8\pi}\bigg[\frac{S_3^2S_1 k^2}{\vert \omega \vert ^6(1- \frac{1}{\omega^2}S_1)(1- \frac{1}{\omega_*^2}S_1) }  \bigg \lbrace \frac{1}{\omega} - \frac{1}{\omega_*}  \bigg \rbrace  + \frac{S_3^2 k^2}{\vert \omega \vert ^4 } \bigg \lbrace \frac{1}{\omega_*(1- \frac{1}{\omega_*^2}S_1)} -  \frac{1}{\omega(1- \frac{1}{\omega^2}S_1)} \bigg \rbrace \bigg]\vert E_{1x} \vert ^2
\end{split}
\label{JdotE}
\end{equation}
where $ \omega_* $ is the complex conjugate of $ \omega$. Similarily, the expression for time derivative of total field energy Eq.(\ref{conserhomo}) can be written as,
\begin{equation}
\begin{split}
\frac{\partial}{\partial t}\bigg(\frac{E^2 + B^2}{8\pi}\bigg) &= \frac{\partial}{\partial t} \bigg[\frac{1}{8\pi	} (E_{1x} \cdot E_{1x}^* + E_{1y} \cdot E_{1y}^*+B_{1z} \cdot B_{1z}^*) \bigg]\\
& = \frac{\partial}{\partial t} \bigg[ \bigg \lbrace 1 +  \frac{k^2S_3^2}{\vert \omega \vert^6(1- \frac{1}{\omega^2}S_1)(1- \frac{1}{\omega_*^2}S_1) } +\frac{k^2}{\vert \omega \vert^2 } \bigg \rbrace \bigg]\vert E_{1x} \vert ^2
\end{split}
\label{EsqBsq}
\end{equation}
The combination of the above two equation Eq.(\ref{JdotE}) and Eq.(\ref{EsqBsq}) gives the energy conservation theorem as 
\begin{equation}
\begin{split}
\frac{\partial}{\partial t}\bigg(\frac{E^2 + B^2}{8\pi}\bigg)+ \vec{J} \cdot \vec{E} &=  \frac{\partial}{\partial t} \bigg[ \bigg \lbrace 1 +  \frac{k^2S_3^2}{\vert \omega \vert^6(1- \frac{1}{\omega^2}S_1)(1- \frac{1}{\omega_*^2}S_1) } +\frac{k^2}{\vert \omega \vert^2 } \bigg \rbrace \vert E_{1x} \vert ^2\bigg]\\
&+\frac{\iota}{8\pi}\bigg[S_3^2 k^2 \bigg\lbrace \frac{1}{\omega^5(1- \frac{1}{\omega^2}S_1)} - \frac{1}{\omega_*^5(1- \frac{1}{\omega_*^2}S_1)} \bigg \rbrace + S_4 k^2 \bigg \lbrace \frac{1}{\omega^3} - \frac{1}{\omega_*^3}\bigg \rbrace \bigg]\vert E_{1x} \vert ^2 \\
& + \frac{\iota}{8\pi}\bigg[ S_2 \bigg \lbrace \frac{1}{\omega} - \frac{1}{\omega_*}  \bigg \rbrace \bigg]\vert E_{1x} \vert ^2\\
& +  \frac{\iota}{8\pi}\bigg[\frac{S_3^2S_1 k^2}{\vert \omega \vert ^6(1- \frac{1}{\omega^2}S_1)(1- \frac{1}{\omega_*^2}S_1) }  \bigg \lbrace \frac{1}{\omega} - \frac{1}{\omega_*}  \bigg \rbrace  + \frac{S_3^2 k^2}{\vert \omega \vert ^4 } \bigg \lbrace \frac{1}{\omega_*(1- \frac{1}{\omega_*^2}S_1)} -  \frac{1}{\omega(1- \frac{1}{\omega^2}S_1)} \bigg \rbrace \bigg]\vert E_{1x} \vert ^2=0
\end{split}
\label{Econservfinal}
\end{equation}
It has been shown by  Califano {\it{et al.}} \citep{PhysRevE.58.7837}  that for the geometry considered by us the growth rate is purely imaginary
 We choose to  analyse the Eq.(\ref{Econservfinal}) here for purely growing mode. The terms containing $ S_2 $ and $ S_4 $ can be collected and simplified as 
\begin{equation}
\frac{\iota}{8\pi}\bigg[ S_2 \bigg \lbrace \frac{1}{\omega} - \frac{1}{\omega_*}  \bigg \rbrace  +S_4 k^2 \bigg \lbrace \frac{1}{\omega^3} - \frac{1}{\omega_*^3}\bigg \rbrace \bigg]\vert E_{1x} \vert ^2 = \frac{2\Gamma}{8\pi} \bigg[ \frac{S_2}{ \Gamma^2} - \frac{S_4}{ \Gamma ^4} \bigg] \vert E_{1x} \vert ^2 
\label{s2s4term}
\end{equation}
where   $\omega = \iota \Gamma $ is the growth rate. The  $ S_2 $  term  i.e. $ \frac{\partial }{\partial t}  [ \frac{1}{8\pi} S_2\vert E_{1x} \vert ^2  ] $ merely 
represents the  dressing of  the electric field energy $ E_{1x}^2 $ by the plasma  arising from  the  $ \vec{J} \cdot \vec{E} $ contribution. 
It should be noted that the  term containing $ S_4 $ has the negative sign  and we would see that this term is in fact   responsible for driving the system unstable. 
This  is also consistent with the description provided by   Califano {\em{et al.}} \citep{PhysRevE.58.7837}. 
Now the terms containing the $ S_3 $ in  Eq. (\ref{Econservfinal}) involve three terms which we denote by  $ T_1 $, $ T_2 $ and $ T_3 $ below 
\begin{equation}
\begin{split}
T_1 = \frac{\iota}{8\pi}\bigg[S_3^2 k^2 \bigg\lbrace \frac{1}{\omega^5(1- \frac{1}{\omega^2}S_1)} - \frac{1}{\omega_*^5(1- \frac{1}{\omega_*^2}S_1)} \bigg \rbrace  \bigg]\vert E_{1x} \vert ^2 = \frac{2\Gamma}{8\pi}S_3^2 k^2 \frac{\Gamma^2+S_1}{\Gamma^4(\Gamma^4 + 2S_1 + S_1^2 )}\vert E_{1x} \vert ^2 
\end{split}
\label{s3T1}
\end{equation}
\begin{equation}
\begin{split}
T_2=\frac{\iota}{8\pi}\bigg[\frac{S_3^2S_1 k^2}{\vert \omega \vert ^6(1- \frac{1}{\omega^2}S_1)(1- \frac{1}{\omega_*^2}S_1) }  \bigg \lbrace \frac{1}{\omega} - \frac{1}{\omega_*}  \bigg \rbrace   \bigg]\vert E_{1x} \vert ^2 = \frac{2\Gamma}{8\pi}S_3^2 S_1 k^2 \frac{\Gamma^2+S_1}{\Gamma^4(\Gamma^4 + 2S_1 + S_1^2 )}\vert E_{1x} \vert ^2 
\end{split}
\label{s3T2}
\end{equation}
\begin{equation}
\begin{split}
T_3=\frac{\iota}{8\pi}\bigg[ \frac{S_3^2 k^2}{\vert \omega \vert ^4 } \bigg \lbrace \frac{1}{\omega_*(1- \frac{1}{\omega_*^2}S_1)} -  \frac{1}{\omega(1- \frac{1}{\omega^2}S_1)} \bigg \rbrace \bigg]\vert E_{1x} \vert ^2 = -\frac{2\Gamma}{8\pi}S_3^2  k^2 \frac{\Gamma^2+S_1}{\Gamma^4(\Gamma^4 + 2S_1 + S_1^2 )}\vert E_{1x} \vert ^2 
\end{split}
\label{s3T3}
\end{equation}
Adding all these terms  with contribution from $ S_3 $ a simplified expression can be obtained as 
\begin{equation}
\begin{split}
T_1 + T_2 + T_3= \frac{2\Gamma}{8\pi} \frac{S_1 S_3^2 k^2}{\Gamma^4(\Gamma^4 + 2S_1 + S_1^2 )}\vert E_{1x} \vert ^2 = \frac{\partial }{\partial t} \bigg [ \frac{1}{8\pi} S_1\vert E_{1y} \vert ^2  \bigg]
\end{split}
\label{T1T2T3}
\end{equation}
This again shows that the coefficient of  $ E_{1y } ^2$ gets modified by the plasma response. 
The complete energy conservation equation can then be written  as,
\begin{equation}
\begin{split}
\frac{1}{8\pi} \frac{\partial}{\partial t} \bigg[ \bigg( 1 + \frac{S_2}{\Gamma^2}\bigg)\vert E_{1x} \vert ^2 + \bigg(1+S_1\bigg)\frac{S_3^2 k^2}{\Gamma^4(\Gamma^4 + 2S_1 + S_1^2 )}\vert E_{1x} \vert ^2 + \frac{k^2}{\Gamma^2}\vert E_{1x} \vert ^2 -\frac{S_4}{\Gamma^2} k^2\vert E_{1x} \vert ^2\bigg] =  \frac{1}{8\pi} \frac{\partial}{\partial t}  
\bigg[ \chi \bigg]=0
\end{split}
\label{T1T2T3}
\end{equation}
From this expression it is clear that the dynamics should be so as to conserve the content indicated by $\chi$ within the square bracket . While the first three terms  
of $\chi$ are positive definite the fourth and the last term with the coefficient of $S_4$ is negative. This permits the possibility of growth in $\vert E_{1x} \vert^2$ even while 
$\chi$ is maintained as constant. It is, therefore, clear that the destabilizing term for this instability arises through $S_4$. On the other hand $S_3$ plays a 
stabilizing role.

\section{Summary}

We have shown through energy principle arguments that the excitation of electromagnetic instability in a 
2-D counterstreaming beam plasma system is possibile which leads to current separation and
magnetic field generation. The term responsible for the instability has been identified as $S_4 = \sum_{\alpha} \frac{n_{0 \alpha} v_{0 x \alpha}^2}{n_0 \gamma_{0 \alpha}}$.  
The terms $S_3$ which measures the asymmetry of the flow between the beam and background plasma plays the role of stabilization 
for this particular mode. It is interesting to note that $S_3$ and $S_4$ change their roles for destabilizing the medium 
when the beam and background plasma having a finite transverse extent such that the Poynting flux becomes relevant. 
This has been shown in a recent 
 submission \cite{Das2017}.

\bibliographystyle{ieeetr}  

\bibliography{finite_arxiv}

\end{document}